%% file: kino2015.tex
\documentclass[12pt]{emulateapj}

\usepackage[dvips]{color}

\shorttitle{M87 jet base}
\shortauthors{Kino et al.}

\begin{document}

\title{Magnetization degree at the jet base of M87
derived from the event horizon telescope data:
Testing magnetically driven jet paradigm}
\author{\thanks{Last update: XX, 2015}
M. Kino\altaffilmark{1}, 
F. Takahara\altaffilmark{2},
K. Hada\altaffilmark{3},  
K. Akiyama\altaffilmark{3},
H. Nagai\altaffilmark{3},
B.W. Sohn\altaffilmark{1}
}

\altaffiltext{1}{
Korea Astronomy and Space Science Institute, 
776 Daedeokdae-ro, Yuseong-gu, 
Daejeon 305-348, Republic of Korea} 
\email{kino@kasi.re.kr}
\altaffiltext{2}{Department of Earth and Space Science,
Osaka University, Toyonaka 560-0043, Japan}
\altaffiltext{3}{National Astronomical Observatory of Japan
                 2-21-1 Osawa, Mitaka, Tokyo, 181-8588, Japan}


\begin{abstract}

We explore the degree of magnetization at the jet base of M87 
by using the observational data  of
the event horizon telescope (EHT)  at 230~GHz
obtained by Doeleman et al.
By utilizing the method in Kino et al.,
we derive the energy densities of magnetic fields ($U_{B}$) and
electrons and positrons ($U_{\pm}$) 
in the compact region detected by EHT (EHT-region) 
with its full-width-half-maximum size 
$40~{\rm \mu as}$.
First, we assume that
an optically-thick region 
for synchrotron self absorption (SSA)
exists in the EHT-region. 
Then, we find that the SSA-thick region
should not be too large
not to overproduce the Poynting power at the EHT-region.
The allowed ranges of the angular size and the magnetic field strength 
of the SSA-thick region are
$21~{\rm \mu as} \le \theta_{\rm thick}\le 26.3~{\rm \mu as}$ and
$50~{\rm G}\le B_{\rm tot}\le 124~{\rm G}$, respectively.
Correspondingly $U_{B}\gg U_{\pm} $ is realized in this case.
We further examine the composition of plasma and energy density of protons
by utilizing the Faraday rotation measurement ($RM$) 
at 230~GHz obtained by Kuo et al.
Then, we find that  $U_{B}\gg U_{\pm}+U_{p} $ still holds in the 
SSA-thick region.
Second, we examine the case when  EHT-region is  fully SSA-thin.  
Then we find that $U_{B}\gg U_{\pm}$ still holds 
unless protons are relativistic.
Thus, we conclude that magnetically driven jet scenario in M87
is viable in terms of  energetics close to ISCO scale
unless the EHT-region is fully SSA-thin and  
relativistic protons dominated.

\end{abstract}


\section{Introduction}
\label{sec:intro}

Elucidating the formation mechanism of relativistic jets
in active galactic nuclei (AGNs) is one of the longstanding
challenges in astrophysics. 
Although magnetically driven jet and wind models 
are widely discussed in the literatures 
(e.g., 
Okamoto 1974;
Blandford \& Znajek 1977;
Blandford \& Payne 1982;
Chiueh et al. 1991;
Li et al. 1992;
Uchida 1997;
Okamoto 1999;
Koide et al. 2002;
Tomimatsu \& Takahashi 2003;
Vlahakis \& Konigl 2003;
McKinney and Gammie 2004;
Krolik et al. 2005;
McKinney 2006; 
Komissarov et al. 2007;
Komissarov et al. 2009;
Tchekhovskoy et al. 2011;
Toma \& Takahara 2013;
Nakamura and Asada 2013;
McKinney et al. 2013),
the actual value of the strength of magnetic field ($B$)
at the base of the jet 
is still an open problem.
In order to test magnetic jet paradigm,
it is most essential to clarify the energy density of 
magnetic fields ($U_{B}\equiv B_{\rm tot}^{2}/8\pi$)
and that of particles at the upstream end of the jet
where $B_{\rm tot}$ is the strength of total magnetic fields.

Recently, short-millimeter radio observations  at 1.3~mm
(equivalent to the frequency 230~GHz) 
have been performed against 
the nearby giant radio galaxy M87.
M87 is located at 
a distance of $D_{\rm L}=16.7~{\rm Mpc}$ (Jordan et al. 2005; Blakeslee et al. 2009), 
hosts one of the 
most massive super massive black hole
$M_{\bullet}=(3-6)\times 10^{9}~M_{\odot}$ 
(e.g., 
Macchetto et al. 1997; 
Gebhardt and Thomas 2009; 
Walsh et al. 2013)
and thus M87 is known 
as the best target for studying the upstream end of the jet
(e.g., Junor et al. 1999; Hada et al. 2011).
The Schwarzschild radius is
$R_{s}\equiv 2GM_{\bullet}/c^{2}
\approx 2 \times 10^{15}~{\rm cm}$ 
for the central black hole with
$M_{\bullet}=6\times 10^{9}~M_{\odot}$
where 
$G$ is the gravitational constant and 
$c$ is the speed of light.
This corresponds to the angular size
of $\sim 7~\mu as$.
Hereafter, we set this mass as the fiducial one.
The Event Horizon Telescope (EHT)
composed of stations in Hawaii and the western United States
 has detected a compact region at the base of the M87 jet
 at 230~GHz with its size $40~\mu$as
(Doeleman et al. 2012).
Furthermore, Kuo et al. (2014) obtained the 
first constraint on  the Faraday rotation measure ($RM$)
for M87 using the 
submillimeter array (SMA) at 230~GHz.

Short milli-meter VLBI observations of EHT
at 230~GHz (equivalent to 1.3~mm) 
is crucially  beneficial
in order to minimize the blending effect of 
 sub-structures below the spatial resolutions 
 of telescopes. 
Historically, single dish observations of AGN jets
at centi-meter  waveband (with arc-minute spatial resolution)
revealed that their spectra are flat at cm waveband (Owen et al. 1978).
Marscher (1977) suggested the importance of VLBI observations 
for distinguishing various possible explanations 
for the observed flatness.
Cotton et al. (1980) 
conducted VLBI observations at cm waveband
and found that the flat spectrum results from 
a blending effect of sub-structures with milli-arcsecond (mas) scale.
This was a significant forward step.
However, subsequent VLBI observations have revealed that
such mas scale components still have sub-structures 
when observing them at higher spatial resolution
(i.e., shorter wavelength).
This is a vicious-circle between 
telescopes' spatial-resolutions and 
sizes of sub-structures.
In the case of M87,
we finally start to overcome this problem
since the spatial-resolution of EHT 
almost reaches  one of the fundamental scales,
i.e., ISCO (Innermost Stable Circular Orbit) scale (Doeleman et al. 2012). 
Hence, in this work, 
we will assume the ISCO radius ($R_{\rm ISCO}$)
as the minimum size of  the jet nozzle.

Motivated by the significant observational progresses by EHT, 
we explore  the magnetization degree ($U_{\pm}/U_{B}$)
 in the core of M87 seen at 230~GHz.
We note that Doeleman et al. (2012) did not 
derive $B_{\rm tot}$ and $U_{\pm}/U_{B}$,  and 
we will estimate them at the EHT-region for the first time.
In the theoretical point of view,
we  have developed the methodology for 
the estimation of $U_{\pm}/U_{B}$ 
and $B_{\rm tot}$
in  Kino et al. (2014) (hereafter K14),
and it is also  applicable to 230~GHz.
In K14, we estimated 
$U_{\pm}/U_{B}$ 
and $B_{\rm tot}$
in the radio core at 43~GHz
with $\theta_{\rm FWHM}=110~{\rm \mu as}$ and
$0.7$~Jy.
We obtained the tight constraint of  field strength 
($1~{\rm G}\le B_{\rm tot} \le 15~{\rm G}$), but
the resultant energetics are consistent with
either the $U_{\pm}$-dominated or $U_{B}$-dominated  
($1\times 10^{-5}\le U_{\pm}/U_{B} \le 6\times 10^{2}$). 
The radio core at 230~GHz
with $\theta_{\rm FWHM}=40~{\rm \mu as}$ 
directly corresponds to the upstream end of 
the M87 jet 
and then it would give tightest constraints
for testing the magnetic jet paradigm.
The goal of this work is applying
the method of K14 to  the EHT-detected region 
and exploring its properties.

Opacity of EHT-region against SSA is critically
important. We should emphasize that
it is not clear that EHT-region is SSA-thick
or not  because short-mm VLBI observations
are conducted only at 230~GHz and therefore 
it is not possible to obtain spectral informations at the moment.
Intriguingly,
Rioja \& Dodson (2011) detect the core shift  between
43 and 86~GHz (in Figure~5 in their paper), 
which means that the radio core 
at 86~GHz contains the SSA-thick region.
With the aid of interferometry observations,
we can also infer the turnover frequency.
The fluxes measured by 
IRAM at 89~GHz (Despringre et al. 1996)
and SMA at 230~GHz (Tan et al. 2008)
also  seem to indicate that
the radio core is SSA-thick above 89~GHz
although the data are not obtained simultaneously
(see also Abdo et al. 2009).
The sub-mm spectrum obtained by ALMA 
also shows the spectral break 
above $\sim 100~{\rm GHz}$
(Doi et al. 2013).
Therefore, we may infer that 
SSA turnover frequency 
for the EHT-region is 
above $\sim 100~{\rm GHz}$.
As a working hypothesis, 
we firstly assume that 
the EHT-region includes the SSA-thick region
and apply the method of K14.
We will also discuss the fully SSA-thin  case in \S 6.

The layout of this paper is as follows.
In \S 2, we briefly review the method of K14.
In \S 3, we apply the method to the EHT-region.
In \S 4, the resultant $U_{\pm}/U_{B}$ and $B_{\rm tot}$ are presented.
In \S 5, we further discuss constraints on the proton component.
In \S 6, we discuss the fully SSA-thin case.
In \S 7, we summarize the results and
give account of important future work to be pursued. 
In this work, we define the radio spectral index $\alpha$
as $S_{\nu}\propto \nu^{-\alpha}$.

\section{Method}

Following K14, here we briefly
review the method for constraining
magnetic field and relativistic electrons
in radio cores.

\subsection{Basic assumptions}

First of all, we show main assumptions in this work.

\begin{itemize}

\item
We assume that the emission region is spherical
with its radius $R$ which is defined as
$2R=\theta_{\rm obs}D_{A}$
where 
$\theta_{\rm obs}$,
$D_{A}=D_{L}/(1+z)^{2}$ and 
$D_{L}$ are
the observed angular diameter of the emission region, 
the angular diameter distance and the luminosity distance, respectively.
This is justified by the following observational suggestion.
In the EHT observation of M87 in 2012,
Akiyama et al. (2015) measures
the closure phase of M87 among the three stations
(SMA, CARMA, and SMT).
The closure phase is the sum of visibility phases
on a triangle of three stations 
(e.g., Thompson et al. 2001;
Lu et al. 2012).
Akiyama et al. (2015) shows that 
the measured closure phases are close to zero ($\lesssim \pm 20^\circ$) 
for the structure detected in Doeleman et al. (2012), which is naturally explained by
a symmetric emission region and disfavors significantly asymmetric one.

\item
We do not include the GR effect for simplicity.
The full GR ray-tracing and radiative transfer
may be essential for reproducing detailed shape of 
black hole shadows
(e.g., 
Falcke et al. 2000;
Takahashi et al 2004; 
Broderick \& Loeb 2009;
Nagakura \& Takahashi 2010;
Dexter et al. 2012;
Lu et al. 2014).
However, 
current EHT  can only detect flux from a bright 
region via visibility amplitude
and spatial structure can be 
constrained only by closure-phases
(e.g., Doeleman et al. 2009).
Although
the predicted black hole shadow images in details
seem diverse, the size of the bright region is roughly
comparable to ISCO scale
(e.g., Fish et al. 2013 for review).
Therefore, we do not include the GR effect 
but explore a fairly wide allowed range for the 
bright region size $\theta_{\rm thick}$, i.e.,
from  $\sim R_{\rm ISCO}$ to  $\sim 2R_{\rm ISCO}$
(see sec. 5).

\end{itemize}

\subsection{General consideration}

Given the SSA turnover frequency ($\nu_{\rm ssa}$) and 
the angular diameter size of the emission region at $\nu_{\rm ssa}$,
one can uniquely determine $B_{\rm tot}$ and $K_{\pm}$ 
where $K_{\pm}$ is the normalization factor of relativistic (non-thermal)
electrons and positrons
(e.g., 
Kellerman \& Poliny-Toth 1969; 
Burbidge et al. 1974;
Jones et al. 1974a, 1974b;
Blandford \& Rees 1978;
Marscher 1987).
Recently, K14 points out that
the observing frequency is 
identical to $\nu_{\rm ssa}$
when we can identify the SSA-thick
surface at observing frequency.

As a first step, we assume that 
the EHT-region is a  one-zone sphere 
with isotropic magnetic field ($B_{\rm tot}$)
and 
particle distributions in the present work.
Locally, we denote 
($B_{\perp,\rm local}= B_{\rm tot} \sin\alpha$) 
as the magnetic field strength
perpendicular to the direction of  electron motion 
(Ginzburg \& Syrovatskii 1965, hereafter GS65)
where $\alpha$ is 
the pitch angle
between the vectors
of electron velocity and the magnetic field
(e.g., Rybicki \& Lightman 1979).
Then, we can obtain pitch-angle averaged $B_{\perp,\rm local}$ 
defined as  $B_{\perp}$ as follows:
\begin{eqnarray}\label{eq:Btot-define}
B_{\rm tot}^{2}=\frac{3}{2}B_{\perp}^{2}
\end{eqnarray}
since 
$B_{\perp}^{2}=
(1/4\pi)
\int B_{\rm tot}^{2}\sin^{2}\alpha d\Omega=2B_{\rm tot}^{2}/3$.
(This is a slightly different definition of $B_{\perp}$  in K14.
The corresponding slight changes of numerical factors 
are summarized in Appendix.)
Since we assume isotropic field, hereafter
we choose $B_{\perp}$ direction to the line of sight (LOS).

The number density distribution of relativistic electrons
and positrons
$n_{\pm}(\epsilon_{\pm})$
is defined as (e.g., Eq.3.26 in GS65)
\begin{eqnarray}
n_{\pm}(\epsilon_{\pm})d\epsilon_{\pm} &=& K_{\pm}\epsilon_{\pm}^{-p} d\epsilon_{\pm} \quad
(\epsilon_{\pm,{\rm min}} \le \epsilon_{\pm} \le \epsilon_{\pm,{\rm max}}) ,  
\end{eqnarray}
where 
$\epsilon_{\pm}=\gamma_{\pm}m_{e}c^{2}$,
$p=2\alpha+1$,
$\epsilon_{\pm,{\rm min}}=\gamma_{\pm,{\rm min}}m_{e}c^{2}$, and
$\epsilon_{\pm,{\rm max}}=\gamma_{\pm,{\rm max}}m_{e}c^{2}$ are
the electron energy,
the spectral index,
minimum energy, and
maximum energy of 
relativistic (non-thermal) electrons and positrons, respectively.
Although electrons and positrons may have 
different heating/accerelation process in $e^{-}/e^{+}/p$
mixed plasma (e.g., Hoshino and Arons 1991),
here we assume that  minimum energies of electrons 
and positrons are same for simplicity.
By evaluating the emission at the 
synchrotron self-absorption frequency,
we  obtain 
\begin{eqnarray}\label{eq:B}
B_{\perp}&=& b(p)
\left(\frac{\nu_{\rm ssa, obs}}{1~{\rm GHz}}\right)^{5}
\left(\frac{\theta_{\rm obs}}{1~{\rm mas}}\right)^{4}
\left(\frac{S_{\nu_{\rm ssa},\rm obs}}{1~{\rm Jy}}\right)^{-2} \nonumber \\
&\times& \left(\frac{\delta}{1+z}\right)    ,
\end{eqnarray}
where $b(p)$ is tabulated in Marscher (1983), Hirotani (2005), and K14.
The term $K_{\pm}$
is given by
\begin{eqnarray}
K_{\pm} &=& k(p)
\left(\frac{D_{\rm A}}{1~{\rm Gpc}}\right)^{-1}
\left(\frac{\nu_{\rm ssa,obs}}{1~{\rm GHz}}\right)^{-2p-3}
\left(\frac{\theta_{\rm obs}}{1~{\rm mas}}\right)^{-2p-5} \nonumber \\
&\times&
\left(\frac{S_{\nu_{\rm ssa},\rm obs}}{1~{\rm Jy}}\right)^{p+2}
\left(\frac{\delta}{1+z}\right)^{-p-3},
\end{eqnarray}
where $k(p)$  is tabulated in K14.
The cgs units of $K_{\pm}$ and $k(p)$
depend on $p$:
erg$^{p-1}$cm$^{-3}$.
It is useful to show the explicit expression 
of the ratio $U_{\pm}/U_{B}$ as follows: 
\begin{eqnarray}\label{eq:ueub}
\frac{U_{\pm}}{U_{B}}&=& 
\frac{16\pi}{3b^{2}(p)} 
\frac{k(p)\epsilon_{\pm,\rm min}^{-p+2}}{(p-2)} 
\left(\frac{D_{\rm A}}{1~{\rm Gpc}}\right)^{-1}
\left(\frac{\nu_{\rm ssa,obs}}{1~{\rm GHz}}\right)^{-2p-13}
\nonumber \\
&\times&
\left(\frac{\theta_{\rm obs}}{1~{\rm mas}}\right)^{-2p-13} 
\left(\frac{S_{\nu_{\rm ssa},\rm obs}}{1~{\rm Jy}}\right)^{p+6}
\left(\frac{\delta}{1+z}\right)^{-p-5}  \nonumber \\
&& ({\rm for} \quad p>2)  .
\end{eqnarray}

From this, 
we find that $\nu_{\rm ssa,obs}$ and
$\theta_{\rm obs}$ have the same dependence on $p$.
Using this relation, we can estimate $U_{\pm}/U_{B}$ 
without minimum energy (equipartition $B$ field) assumption.
It is clear that 
the measurement of $\theta_{\rm obs}$ is crucial
for determining $U_{\pm}/U_{B}$.

We further impose two general constraint conditions. 

\begin{enumerate}

\item

Time-averaged total  power of the jet ($L_{\rm jet}$)
estimated by jet dynamics 
at large-scale  should not be exceeded by
the one at the jet base
\begin{eqnarray}\label{eq:energetics1}
L_{\rm jet} &\ge&  
\max
\left[L_{\rm poy}, L_{\pm}\right] ,  \nonumber \\ 
L_{\pm} &=& \frac{4\pi}{3} \Gamma^{2} \beta R^{2} c U_{\pm}, \nonumber \\
L_{\rm poy} &=& \frac{4\pi}{3} \Gamma^{2} \beta R^{2} c U_{B} ,
\end{eqnarray}
where
$L_{\pm}$, 
$L_{\rm poy}$, 
$\Gamma$, and 
$\beta c$
are, respectively,
electron/positron kinetic power,
Poynting power,
bulk Lorentz factor, and
bulk speed
of the jet at the EHT region.
Note that 
$U_{B}$, $U_{\pm}$, $R$ are 
directly constrained by VLBI observations.

\item
The minimum Lorentz factor of relativistic 
electrons and positrons ($\gamma_{\pm,\rm min}$)
should be smaller than the ones radiating the observed synchrotron
emission ($\nu_{\rm syn,obs}$), for example 230~GHz.
Otherwise, we would not be able to observe synchrotron emission
at the corresponding frequency.
This is generally given by 
\begin{eqnarray}\label{eq:nu_syn}
\nu_{\rm syn,obs} \ge   
1.2\times 10^{6}B _{\perp}\gamma_{\pm,\rm min}^{2}
\frac{\delta}{1+z}   .
\end{eqnarray}
These relations significantly constrain 
on the allowed values of $\gamma_{\pm,\rm min}$
and $B_{\rm tot}$.

\end{enumerate}

In the next section, 
we will add another constraint condition 
(i.e., minimums size limit).

\section{Application to the EHT-region}

Here we apply the method to the EHT-region
in M87.

\subsection{On basic physical quantities}

Here we list the basic physical quantities 
of the M87 jet.

\begin{itemize}

\item

The total jet power 
$L_{\rm jet}$ can be estimated
by considering jet dynamics at well-studied
bright knots (such as knots A, D and HST-1) located at kpc scale
(e.g., 
Bicknell \& Begelman 1996;
Owen et al. 2000; 
Stawarz et al. 2006).
Based on the literatures on these  studies, here we adopt
\begin{eqnarray}
1\times 10^{44}~{\rm erg~s^{-1}} \le L_{\rm jet}\le 5\times 10^{44}~{\rm erg~s^{-1}} , 
\end{eqnarray}
(see also Rieger \& Aharonian 2012 for review).
We note that Young et al. (2002) indicates
$L_{\rm jet}\sim  3 \times 10^{42}~{\rm erg~s^{-1}}$
based on the X-ray bubble structure
which is significantly smaller than the aforementioned
estimate.　
The smallness of  $L_{\rm jet}$ estimated by
Young et al. (2002) could be attributed 
to a combination of intermittency 
of the jet and an averaging of $L_{\rm jet}$
on a long time scale of X-ray cavity age. 
In this work, we do not 
utilize this small $L_{\rm jet}$.

\item
We would  assume that 
the bulk speed of the jet is 
in non-relativistic regime
at the jet at the EHT region
since both theory and observations 
currently tend to indicate slow and gradual acceleration
so that the flow reaches  the relativistic speed
around $10^{3-4}~R_{s}$
(McKinney 2006; 
Asada \& Nakamura 2014;
Hada et al. 2014).
The brightness temperature of the 230~GHz radio core
is below the critical temperature $\sim 10^{11}~{\rm K}$ 
limited by inverse-Compton catastrophe process
(Kellermann \& Pauliny-Toth 1969).
When the 230~GHz emission originates from the 
SSA-thick plasma, the characteristic 
electron temperature is comparable to
$T_{b}$ (e.g., Loeb \& Waxman 2007)
and $T_{b}$ at 230~GHz is in relativistic regime.
 Therefore, we set 
 \begin{eqnarray}
 \Gamma\beta =c_{\rm sound=}
  \frac{1}{\sqrt{3}}    ,
 \end{eqnarray}
where $c_{\rm sound}$
is the sound speed of relativistic matter.
This  will be used in Eq.~(\ref{eq:energetics1})
as $ \Gamma^{2}\beta =1/\sqrt{3} $.

\item

Last, we summarize three differences between 
this work and  Doeleman et al. (2012)
in terms of the assumptions on basic physical quantities.
In this work, we attempt to reduce assumptions 
and treat the EHT-region in a more general way.
(1) 
In Doeleman et al. (2012), they assume that
the EHT-region size is identical to the ISCO size itself
which reflects the degree of the black hole spin.
In this work,  we do not use this assumption.
(2)
Doeleman et al. (2012) seems to focus on 
the SSA-thin case.
In this work, we will investigate both 
SSA-thick and SSA-thin cases.
(3)
Doeleman et al. (2012) seems to
assume $\theta_{\rm FWHM}$ as 
the physical size of the EHT-region.
In this work, we take into account a
deviation factor between 
 $\theta_{\rm FWHM}$ and its 
 physical size (e.g., Marscher 1983).

\end{itemize}

\subsection{Difficulties for SSA-thick one-zone model}

First,
we estimate the magnetic field strength
in the EHT region by assuming that 
all of the EHT-region with $\theta_{\rm FWHM}=40~{\rm \mu as}$
is fully SSA-thick. 
The field strength of EHT-region is estimated as
\begin{eqnarray}
B_{\rm tot}&= & 3.4\times 10^{2}~{\rm G}~
\left(\frac{\nu_{\rm ssa, obs}}{230~{\rm GHz}}\right)^{5} \nonumber \\
&\times&
\left(\frac{\theta_{\rm obs}}{72~{\rm \mu as}}\right)^{4}
\left(\frac{S_{\nu_{\rm ssa},\rm obs}}{1.0~{\rm Jy}}\right)^{-2}
\left(\frac{\delta}{1+z}\right).
\end{eqnarray}
%
Marscher (1983) pointed out 
VLBI measured $\theta_{\rm FWHM}$ is 
connected with true angular size
$\theta_{\rm obs}$ by the relation
$\theta_{\rm obs}\approx 1.8 \theta_{\rm FWHM}$
for partially resolved sources.
(see also Krichbaum et al.  2006; Loeb and Waxman 2007).
Taking  such deviation into account, we examine the case of
$72~{\rm \mu as}=1.8\times 40~{\rm \mu as}$
for the estimate of $B$-field strength.

What happens with this field strength?

\subsubsection{Too large Poynting Power}

A severe problem arises  
if  $B_{\rm tot}\approx 3\times 10^{2}~{\rm G} $ is realized.
Since we assume nearly isotropic random field which 
can be supported by the low linear polarization degree at 230~Ghz
(Kuo et al. 2014), the corresponding Poynting power is given by 
\begin{eqnarray}
L_{\rm poy}&=&1.5\times 10^{47}~{\rm erg~s^{-1}}\nonumber\\
&\times&
\left(\frac{B_{\rm tot}}{300~{\rm G}}\right)^{2}
\left(\frac{2R}{1.8\times 10^{16}~{\rm cm}}\right)^{2} . 
\end{eqnarray}
Here we adopt
$2R=1.8\times 10^{16}~{\rm cm}=
1.8\times 40~\mu as\times 16.7~{\rm Mpc}$.
When total power  of the jet
(i.e., sum of kinetic and Poynting ones) is conserved
along the jet at a large-scale,
then this is too large
compared with the jet's mean kinetic power inferred
from its large-scale dynamics 
a few$\times 10^{44}~{\rm erg~s^{-1}}$
(e.g., Rieger \& Aharonian 2012 for review)
We emphasize that
a constraint on $B_{\rm tot}$
by $L_{\rm poy}$ is almost model-independent.

If we allow some kind of
fast magnetic reconnection processes 
(e.g., 
Kirk \& Skj{\ae}raasen 2003;
Bessho \& Bhattacharjee 2007;
Takamoto et al. 2012;
Bessho \& Bhattacharjee 2012)
in order to dissipate magnetic fields
at the EHT region,
then fast and large variabilities would be naturally expected.
However, there is no observational support for such variabilities.
Therefore, it seems difficult to realize
too large $B_{\rm tot}$ at EHT region.

\subsubsection{Too fast synchrotron cooling}

Once we obtain a typical value of $B_{\rm tot}$,
then we can estimate a typical synchrotron cooling timescale.
It significantly characterizes observational behavior of
the EHT region.
The synchrotron cooling timescale is correspondingly
\begin{eqnarray}
t_{\pm, \rm syn} \approx 1\times 10^{-2} ~{\rm day}
\left(\frac{B_{\rm tot}}{300~{\rm G}}\right)^{-2}
\left(\frac{\gamma_{\pm}}{10}\right)^{-1}  .
\end{eqnarray}
This is much shorter than day-scale although
the flux at 230~GHz measured by the EHT 
remains constant during subsequent three days 
(see the Supplementary Material of Doeleman et al. 2012).
Then, a difficulty arises due to this short $t_{\pm,\rm syn}$.
The 230~GHz radio emitting electrons are in the
so-called fast cooling regime  (Sari et al. 1998)
in which injected electrons instantaneously 
cool down  by synchrotron cooling.
Hence, 
a slight change/fluctuation of $B$ field strength 
instantaneously 
(on timescale $t_{\pm,\rm syn}$) 
is reflected on the  synchrotron flux at EHT region.
Hence, for realizing the observed constant flux,
 a constant plasma supply of 
$B_{\rm tot}$ and $K_{\pm}$ with very small fluctuation is required
to avoid rapid variability/decrease of synchrotron flux.
On the other hand, when the magnetic fields are not
that large, $t_{\pm,\rm syn}$ can become longer than day scale.
Then, we can avoid rapid variability/decrease of synchrotron flux
without imposing  very small fluctuation of $B_{\rm tot}$ and $K_{\pm}$
in the bulk flow.
Since some fine-tuning of  $B_{\rm tot}$ and $K_{\pm}$ injection may
be able to adjust the observed constant flux density 
at the EHT region,
the too-fast-cooling problem may be less severe than
the aforementioned problem on too-large-$L_{\rm poy}$.
But it is natural to suppose that smaller $B_{\rm tot}$ realizes
in the EHT region to avoid fine tuning of injection quantities.

\subsection{Two-zone model}

\subsubsection{Basic idea}

\begin{figure} 
\includegraphics[width=8cm]{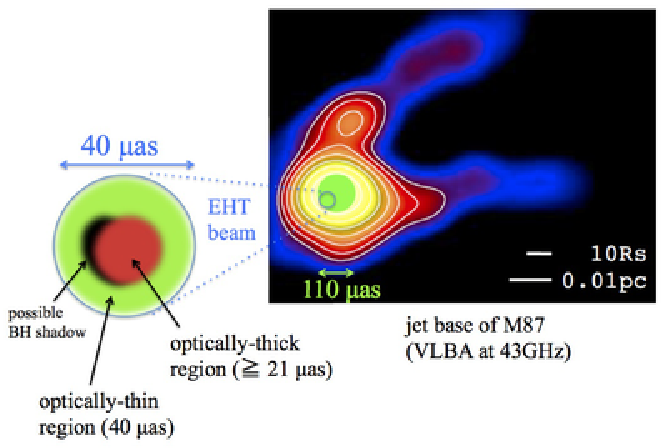}
\caption
{Illustration of the jet base of M87
down to the EHT-region scale.
The right panel shows the actual image of 
M87 with VLBA at 43~GHz adopted from Hada et al. (2013).
The yellow-green circle shows the one-zone
region with its diameter $110~{\rm \mu as}$ which
is investigated in K14. 
The EHT-region  detected by Doeleman et al. (2012)
is shown as the blue circle. 
Since Hada et al. (2011)
indicate that
the central engine of M87 is located at 
$\sim 41~{\rm \mu as}$ eastward of the radio core at 43~GHz, we put the 
the EHT-region around there.
The left panel shows the illustration of internal structure
inside the EHT-region. 
The red-colored region represents
an SSA-thick compact region inside the SSA-thin region.
The black-colored region conceptually shows a 
possible BH shadow image.
According to  the smallness of closure phase reported in  
Akiyama et al. (2015), a certain level of symmetry is 
kept in this picture. }
\label{fig:illustration}
\end{figure}

The difficulty of too-large-$L_{\rm poy}$ can be resolved 
if the EHT-regions is composed of SSA-thick and SSA-thin regions and
the angular-size of SSA-thick region ($\theta_{\rm thick}$)
 is more compact 
 than $\theta_{\rm obs}$, 
 i.e., 
\begin{eqnarray}
\theta_{\rm obs}> \theta_{\rm thick} .
\end{eqnarray}
We show an illustration of our scenario  in Figure~\ref{fig:illustration}.
In this solution, most of the correlated flux density detected by
EHT is attributed to the emission from the SSA-thin region. 
Since $\nu_{\rm ssa}$ of the SSA-thin region is 
by definition smaller than 230~GHz, the magnetic field must be
significantly smaller because the field strength is proportional
to $\nu_{\rm ssa}^{5}$. 
Because of this reason, we regard 
the SSA-thick region as 
the main carrier of the Poynting power.

Here, we assume the ISCO 
radius for non-rotating black hole
($R_{\rm ISCO}
=6GM_{\bullet}/c^{2}=3R_{s}
\equiv \theta_{\rm ISCO}D_{L}$)
as the minimum size of SSA-thick region.
This corresponds to the angular size $21~{\rm \mu as}$.
Indeed, theoretical  works  
(Broderick \& Loeb 2009; Lu et al. 2014) 
comparing EHT observations and jet models 
also indicate the model images with 
short-mm bright region
\footnote{Conventionally, such regions are sometimes
called as hot spots in literatures (e.g., Lu et al. 2014 and 
reference therein)}
 with their size comparable to ISCO.
Therefore, we examine the range of 
SSA-thick region
$\theta_{\rm thick}\ge 21~{\rm \mu as}$.

\subsubsection{Gaussian fitting with two-components}

In Figure~\ref{fig:radplot}, we estimate the 
correlated flux density of this 
SSA-thick region based on the EHT data.
The observed flux data plotted as a function
of baseline length are adopted from Doeleman et al. (2012). 
The black solid curve is the best-fit circular gaussian
model by Doeleman et al. (2012).
The red solid curve is the best-fit model.
The red dashed  and dot-dashed curves represent
the SSA-thick and the SSA-thin components,
respectively.

Below, we explain the details of the gaussian fitting.
To determine the correlated flux density 
for the compact SSA-thick region with
its lower limit
size $\theta_{\rm FWHM}=21~{\rm \mu as}/1.8=11.1~{\rm \mu as}$,
we  conduct the two-component (SSA-thick and thin components)
gaussian fitting to the EHT-data.
First, we obtain the  upper limit of 
the correlated flux density for
the SSA-thick component as $S_{\nu}=0.27~{\rm Jy}$.
Next, we perform the 
the two-component (SSA-thick and thin components)
gaussian fitting by fixing 
$\theta_{\rm FWHM}=21~{\rm \mu as}/1.8=11.1~{\rm \mu as}$ 
and $S_{\nu}=0.27~{\rm Jy}$.
Then, we obtain the corresponding size and flux of the
extended SSA-thin component
$S_{\nu}=0.75~{\rm Jy}$ and
$\theta_{\rm FWHM}=60~{\rm \mu as}$.

\section{Results}

Here, we limit on 
$B_{\rm tot}$, $\theta_{\rm thick}$, $U_{\pm}/U_{B}$, 
in the EHT-region without assuming plasma composition.
The critical value $\gamma_{\pm,\rm min} $ is 
derived by the combination of  
the jet power limit (Eq.~\ref{eq:energetics1}) and
synchrotron emission limit  (Eq.~(\ref{eq:nu_syn})).
By eliminating  $B_{\rm tot}$, 
we obtain 
\begin{eqnarray}\label{eq:gmin_EHT}
\gamma_{\pm,\rm min}&\le &
1.2\times 10^{2}  \nonumber \\
&\times&\left(\frac{2R}{1.8 \times 10^{16}~{\rm cm}}\right)^{1/2}   
\left(\frac{L_{\rm jet}}{5\times 10^{44}~{\rm erg~s^{-1}}}\right)^{-1/4},    \nonumber \\
\end{eqnarray}
where $\nu_{\rm ssa}=230~{\rm GHz}$ is used.
Since $\gamma_{\pm,\rm min}$
has $R$ dependence, larger $R$ allows slightly larger
$\gamma_{\pm,\rm min}$.

In Figure~\ref{fig:synlimit_fiducial}, we show 
the value of $\log (U_{\pm}/U_{B})$
in the allowed ranges of $\gamma_{\pm,\rm min}$ and $B_{\rm tot}$
with 
$L_{\rm jet}=5\times 10^{44}~{\rm erg~s^{-1}}$
and $p=3.0$.
It is essential to note that
the maximum value of $B_{\rm tot}$
is determined by the condition $L_{\rm poy} \le L_{\rm jet}$
while the minimum value of $B_{\rm tot}$ is governed by 
the condition of 
$\theta_{\rm thick}\ge R_{\rm ISCO}/D_{\rm L}\approx 21~\mu$as.
The right side of the allowed region is determined by
the $\nu_{\rm syn,obs}$ limit shown in Eq.~(\ref{eq:nu_syn}).
Note that the maximum value of 
$\theta_{\rm thick}=26.3~\mu{\rm as}$
is smaller than $40~\mu{\rm as}$.
This suggests that the EHT-region has a more compact
SSA-thick component in it.
Interestingly, overall SSA-thick region 
satisfies $U_{B}\gg U_{\pm}$.
If  protons do not contribute to jet energetics,
then this result
 supports the magnetically driven jet scenario.
In Table~\ref{table:synlimit_fiducial}  
we show the  resultant allowed values.
Summing up, we find that
(1) the allowed  $\theta_{\rm thick}$ satisfies
$21~{\rm \mu as}\le \theta_{\rm thick}\le 26.3~{\rm \mu as}$, 
and that
(2)
the allowed fields strength is  
$50~{\rm G}\le B_{\rm tot}\le 124~{\rm G}$.

When we choose  a smaller $L_{\rm j}$, 
the upper limit of 
$\theta_{\rm thick}$ and $B_{\rm tot}$ 
 becomes smaller according to
 Eqs.~(\ref{eq:B}) and (\ref{eq:energetics1}).
When $L_{\rm j} = 1\times 10^{44}~{\rm erg~s^{-1}}$, 
the allowed  $ B_{\rm tot}$  and $\theta_{\rm thick}$ are
$50~{\rm G}\le B_{\rm tot}\le 65~{\rm G}$ and
$21~{\rm \mu as}\le \theta_{\rm thick}\le 22.4~{\rm \mu as}$.
In this case, the allowed regions of 
$ B_{\rm tot}$  and $\theta_{\rm thick}$  are 
very narrow.

We add to note
a short comment on brightness temperature.
The brightness temperature of the SSA-thick region 
can be estimated as
\begin{eqnarray}
T_{b}&=&
\frac{S_{\nu,\rm obs} c^{2}}
{2\pi k \nu_{\rm syn,obs}^{2}(\theta_{\rm thick}/2)^{2}}   \nonumber \\
&\approx& 2\times 10^{10}~K
\left( \frac{S_{\nu,\rm obs}}{0.27~{\rm Jy}}\right)
\left( \frac{\theta_{\rm thick}}{21~{\mu\rm as}}\right)^{-2}  ,
\end{eqnarray}
 where $\nu_{\rm syn,obs}=230~{\rm GHz}$.
 This value is comparable with the $T_{b}$ at 86~GHz
 estimated by Lee (2013).

Last, it is worth to add one thing.
According to 
the equation of state in relativistic temperature regime
(e.g., 
Chandrasekhar 1967;
Kato, Fukue \& Mineshige 1998), 
we obtain 
\footnote{
When magnetic fields are uniform,
the numerical factor at the right-hand side
in Eq.~(\ref{eq:Tblimit})  is smaller than this case because of 
fewer degree of freedom for electrons/positrons
(see Slysh 1992; Tsang and Kirk 2007).}
\begin{eqnarray}\label{eq:Tblimit}
\gamma_{\pm,\rm ssa}\approx
3\frac{kT_{b}}{m_{e}c^{2}} 
\approx
10 \left(\frac{\theta_{\rm thick}}{21~{\rm \mu as}}\right)^{-2}   ,
\end{eqnarray}
 where we use the fact that $\gamma_{\pm,\rm min}m_{e}c^{2}$
 can be identical to the average energy of electrons and positrons
 since $p$ is steeper than $2$.
The obtained $\gamma_{\pm,\rm ssa}$
tends to be smaller than the minimum Lorentz factor obtained by
Eq.~(\ref{eq:gmin_EHT}) by a factor of a few.
While we may use  $\gamma_{\pm,\rm ssa}$
as  $\gamma_{\pm,\rm min}$,
we conservatively use the condition
Eq.~(\ref{eq:gmin_EHT})
 taking some uncertainty of numerical factor 
 in Eq. (\ref{eq:gmin_EHT}) into account.

\section{Constraints on proton component}\label{sec:toy-model}

In \S~\ref{sec:toy-model},
we investigate constraint on the energy density of protons ($U_{p}$) by 
using Faraday $RM$ measured by Kuo et al. (2014).
From the measured $RM$, we will constrain 
the number density of protons ($n_{p}$).
Then,  we examine $U_{p}$.
The degree of proton contribution  in energetics
has a significant influence over  
relativistic jet formation 
(e.g.,  Begelman, Blandford \& Rees 1984;
Reynolds et al. 1996).

\subsection{Further Assumptions}

To discuss the proton contribution, we need to add  some 
further assumptions.
Although the observed radio emissions warrant 
the existence of relativistic $e^{-}/e^{+}$ population, 
it is not clear about the origin of 
relativistic $e^{-}/e^{+}$ which radiate radio emissions at 230~GHz.
There are several possibilities for its origin. 
Relativistic protons may play an important role
for  heating/acceleration of positrons via resonance process
 with relativistic protons in shocked regions
(e.g., Hoshino and Arons 1991),
while  direct $e^{\pm}$ pair injection 
(Iwamoto \& Takahara 2002;
Asano \& Takahara 2009), and/or
relativistic neutron injection
(Toma \& Takahara 2012)
processes may also work at the jet formation regions.
It is beyond the scope of this work
to clarify  the origin of relativistic $e^{-}/e^{+}$ population
and their relation with proton component.
In this section, we simply assume the existence of 
protons  and generally define the average
energy of  these protons as $\epsilon_{p}$.

As mentioned in the Introduction,
Kuo et al. (2014) obtained the first
constraint on RM for M87 using SMA
at 230~GHz.
Although it is not clear 
how much fraction of linearly polarized emission
comes from the EHT-region,
it is worth to extend the method used in the previous sections
by including $RM$ constraint and apply to
the present case of 230~GHz core of M87.
The degree of LP  $\sim 1~\%$ at 230~GHz
detected by Kuo et al (2014)
is significantly smaller than
the value when fully-ordered field 
(i.e., typically $\sim 70~\%$ for SSA-thin case and 
$\sim 16~\%$ for SSA-thick case, see Pacholczyk 1970).
Hence,  the assumption of isotropic $B$-fields in this work 
looks reasonable to some extent.
On the other hand, only ordered
magnetic fields  aligned to the line of sight  
($B_{\rm LOS}$) contribute to the $RM$.
Hereafter, we   conservatively assume
$B_{\rm tot} \ge B_{\rm LOS}$.

\subsection{$RM$ limit}

Here we    introduce a new constraint
using $RM$ observation data.
This $RM$ is important for estimating 
the kinetic power of protons ($L_{p}$)
because $RM$ can constrain the  proton number density.
Generally speaking, an observed rotation measure ($RM_{\rm obs}$)
consists of two parts, i.e., RM by internal (jet) ($RM_{\rm jet}$)
and RM by external (foreground) matter ($RM_{\rm ext}$).
Therefore,  the $RM_{\rm obs}$ can be decomposed into 
\begin{eqnarray}
RM_{\rm obs}= RM_{\rm jet}+RM_{\rm ext} .
\end{eqnarray}
%
Basically, it is difficult to decouple 
$RM_{\rm jet}$ and $RM_{\rm ext} $ and obtain
$RM_{\rm jet}$. However, it may be possible to
discuss an upper limit of $|RM_{\rm jet}|$ 
with some reasonable assumptions.
When the observed RM  ($RM_{\rm obs}$) is 
comparable to  $RM_{\rm ext}$, then  we obtain
\begin{eqnarray}
RM_{\rm obs}\approx RM_{\rm ext}, \quad
| RM_{\rm jet} |  \ll RM_{\rm obs} .
\end{eqnarray}
Indeed,
foreground Faraday screen 
in close vicinity of jets seems to well explain 
observed $RM_{\rm obs}$
 for radio-loud AGNs
(e.g., Zavala \& Taylor 2004) .
The explicit form of $RM_{\rm jet}$
the rotation measure for relativistic plasma is given as
\begin{eqnarray}\label{eq:RMjet}
| RM_{\rm jet} | &=&\frac{e^{3}}{2\pi m_{e}^{2}c^{4}}
\int dl B_{\rm LOS} 
n_{p} \frac{\log \gamma_{\pm,\rm min}}{2\gamma_{\pm,\rm min}^{2}} \nonumber \\
&\le & 5.36\times 10^{3}
B_{\rm tot} n_{p} \frac{\log \gamma_{\pm,\rm min}}{2\gamma_{\pm,\rm min}^{2}}
\left(\frac{R}{10^{16}~{\rm cm}}\right) ~
{\rm rad~m^{-2}} ,  \nonumber \\
\end{eqnarray}
where we set $\int dl \approx 2R$ since the region is assumed as uniform.
From this, we see that Faraday rotation is strongly suppressed
in relativistic plasma
(Jones \& Odell 1977; 
Quataert \& Gruzinov 2000;
Broderick \& McKinney 2010).
Note that RM only include ionic plasma contribution,
and does not include the electron/positron pair plasma.
It is because electron and positron 
have the same mass but have opposite 
(i.e., minus and plus) charges and then the net Faraday
rotation by them is cancelled out.
Qualitatively saying,
the mixture of $e^{\pm}$ pair plasma (i.e., $\eta<1$)
effectively reduce the value of $RM_{\rm jet}$.

Regarding $RM$-limit of M87,
Kuo et al. (2014)  has measured
 $| RM_{\rm obs}|  \approx (3.4-7.5)\times 10^{5}~{\rm rad~m^{-2}}$
and 
they assume $RM_{\rm obs}\approx RM_{\rm ext}$.
Following Kuo et al. (2014), we also assume $RM_{\rm obs}\approx RM_{\rm ext}$.
Then, the $RM$-limit can be written as
\begin{eqnarray}
| RM_{\rm jet} |  \le  1\times 10^{5}~{\rm rad~m^{-2}}  .
\end{eqnarray}

Note that the above constraint only gives
the upper limit of $n_{p}$.
Therefore, the finite value of  $| RM_{\rm jet} | $
does not exclude the plasma composition
of pure $e^{\pm}$ plasma.

In sub-section 5.4, we will
constrain proton contributions
in the case of
$B_{\rm tot}\approx B_{\rm LOS}$
in Eq~(\ref{eq:RMjet}).
At the moment, this is the only case 
which we can deal with within 
this simple framework.

\subsection{Plasma composition and $e^{\pm}/p$-coupling rate}

To further constrain physical properties at the jet base,  
here we introduce the basic
plasma properties and define general notations.
The number densities of 
protons ($n_{p}$) positrons ($n_{+}$), 
and electrons ($n_{-}$) are, respectively, 
defined as follows:
\begin{eqnarray} \label{eq:eta}
n_{p}&\equiv&\eta n_{-}   ,  \nonumber \\
n_{+}&=&      (1-\eta)n_{-}  \quad (0\leq \eta\leq 1)    ,  \nonumber \\
n_{p}&=&n_{e^{-}p}  \approx \frac{\eta}{2-\eta}
\frac{1}{p-1}K_{\pm} \gamma_{\pm,\rm min}^{-p+1}  ,
\end{eqnarray}
where 
$\eta$ is a free parameter 
describing the proton-loading in the jet.
Here we use the charge neutrality condition in the jet.
It is convenient to define further quantities:
\begin{eqnarray} \label{eq:eta}
n_{\pm}&\equiv& n_{-}+n_{+}=(2-\eta)n_{-}, \nonumber \\
n_{e^{-}p}&\equiv&\eta n_{-} =n_{p}  ,
\end{eqnarray}
where 
$n_{\pm}$ and 
$n_{e^{-}p}$ are, the 
number density
of electrons and positrons,  and that of
proton-associated electrons, respectively.
The case of $\eta=0$ corresponds to pure $e^{\pm}$ plasma
while  $\eta=1$ corresponds to the pure $e^{-}/p$ plasma.
Next, it is important to clarify 
energy balances between electrons and protons.
It is useful to introduce
the parameter defining the 
average energy 
ratio between protons and electrons as $\zeta$ as 
\begin{eqnarray}
\epsilon_{\pm}
\equiv \zeta \epsilon_{p}, \quad 
(\frac{m_{e}}{m_{p}}\le \zeta\le 1),
\end{eqnarray}
where
$\epsilon_{\pm}$ is the average energy of 
relativistic $e^{\pm}$.
The case $\zeta=1$ can be realized 
for equipartition between electrons, positrons and protons
via effective $e^{\pm}/p$ coupling
while $\zeta=m_{e}/m_{p}$ means inefficient $e^{\pm}/p$ coupling
for example 
through randomization of bulk kinetic energy of the jet flow
(e.g., Kino et al. 2012 and reference therein). 
Since we focus on the case of $p>2$ suggested 
in M87 (Doi et al. 2013),
relativistic electrons at minimum Lorentz factors characterize the
total energetics.
Here,  $\epsilon_{\pm}\approx \gamma_{\pm, \rm min}m_{e}c^{2}$
can be estimated as $0.5~{\rm MeV}\le \epsilon_{\pm} \le 50~{\rm MeV}$
together with $1\le \gamma_{\pm,\rm min}\le 100$ based on the 
obtained $\gamma_{\pm, \rm min}$.
Then,
the case $\zeta=1$ corresponds to that of non-relativistic protons 
($0.5~{\rm MeV}\le \epsilon_{p} \le 50~{\rm MeV}$)
while
the case $\zeta=m_{e}/m_{p}$ coincides with that of relativistic protons
($1~{\rm GeV}\le \epsilon_{p} \le 100~{\rm GeV}$).

In general, $L$ is decomposed to
\begin{eqnarray}\label{eq:p-def}
L_{\rm jet}&=&          L_{\pm}+L_{p}+L_{\rm poy} , \nonumber \\   
L_{\pm}     &\equiv& L_{-}+L_{+},
\end{eqnarray}
where 
$L_{\pm}$,
$L_{-}$,
$L_{+}$,
$L_{p}$, and
$L_{\rm poy}$ are, the powers of 
the sum of electrons and positrons,
electrons,
positrons,
protons, and
magnetic fields respectively.
For convenience, we define $\eta_{\rm eq}$
for $L_{p}=L_{\pm}$ and it is given by
\begin{eqnarray}
\eta_{\rm eq} &\equiv&\frac{2\zeta}{1+\zeta} , \nonumber \\
L_{p}&>& L_{\pm}  \quad {\rm for} \quad \eta>\eta_{\rm eq} , \nonumber \\
L_{p}&=& L_{\pm}  \quad {\rm for}  \quad \eta=\eta_{\rm eq} , {\rm and}\nonumber \\
L_{p}&<& L_{\pm}  \quad {\rm for}  \quad \eta<\eta_{\rm eq} . 
\end{eqnarray}
%
Since we set
\begin{eqnarray}
U_{\pm}&\approx& \epsilon_{\pm}n_{\pm}, \\ 
U_{p}&\approx& \epsilon_{p}n_{p}=\epsilon_{p}n_{e^{-}p}  ,
\end{eqnarray}
$L_{p}/L_{\pm}=U_{p}/U_{\pm}=\eta/[(2-\eta)\zeta]$ holds.
Finally,
time-averaged total  power of the jet ($L_{\rm jet}$)
can be generalized as follows:
\begin{eqnarray}\label{eq:energetics2}
L_{\rm jet} &\ge&  
\max
\left[L_{\rm poy}, 
\left(1+\frac{\eta}{2-\eta}\frac{1}{\zeta}\right)
L_{\pm}\right] .
\end{eqnarray}
Given the two model parameters
$\eta$ and  $\zeta$, we obtain $U_{p}$.

\subsection{Limits on 
$B_{\rm tot}$, $\theta_{\rm thick}$, $U_{\pm}/U_{B}$, and 
$U_{p}/U_{B}$}

Here, we give limits on 
$B_{\rm tot}$, $\theta_{\rm thick}$, $U_{\pm}/U_{B}$, and 
$U_{p}/U_{B}$ in the EHT-region for $e^{-}/e^{+}/p$ mixed plasma.
As for plasma properties, the following four cases
with proton loaded plasma can be considered,
i.e., 
relativistic protons with $e^{-}/p$-dominated composition,
relativistic protons with $e^{\pm}$-dominated composition,
non-relativistic protons with $e^{-}/p$-dominated composition, and
non-relativistic protons with $e^{\pm}$-dominated composition.

\subsubsection{The case for relativistic protons ($\zeta=m_{e}/m_{p}$)}

Here we consider the case for 
relativistic protons ($\zeta=m_{e}/m_{p})$.
In Figure ~\ref{fig:eta99zetamemp},  
we show a typical example  of 
"$e^{-}/p$-dominated" case with $\eta=0.99$.
In this case, we obtain $\eta_{\rm eq}=1.09\times 10^{-3}$.
Since we consider  "$e^{-}/p$-dominated" composition,
the upper limit of $RM$ significantly
constrains smaller $\gamma_{\pm,\rm min}$ 
according to Eq.~(\ref{eq:RMjet}).
In this case,  $U_{B}\gg U_{\pm}$ still holds 
as smaller $\gamma_{\pm,\rm min}$ region
is excluded by the $RM$ constraint.
In Table~\ref{table:eta99zetamemp}, 
we summarize the resultant allowed 
physical quantities in this case. 
The maximum value of $B_{\rm tot}$
is determined by the condition $L_{\rm poy} \le L_{\rm jet}$
while the minimum value of $B_{\rm tot}$ is governed by 
the condition that 
$\theta_{\rm thick} \ge R_{\rm ISCO}/D_{\rm L}\approx 21~\mu$as.
In the limit of inefficient $e^{\pm}/p$ coupling,
minimum energy of electrons/positrons are  
smaller than that of protons 
by a factor of $m_{e}/m_{p}$ 
(i.e., $\epsilon_{\pm,\rm min}=(m_{e}/m_{p})\epsilon_{p,\rm min}$).
Therefore, $L_{\pm}$ decreases
and $L_{p}$ tends to dominate over $L_{\pm}$.
The energetics constraint in this case is given by
$L_{\rm jet} \ge
\max
\left[L_{\rm poy}, 
\left(1+
\frac{\eta}{2-\eta}\frac{m_{p}}{m_{e}}\right)
L_{\pm}\right] $.

In the case of  $e^{\pm}$-dominated composition with
smaller $\eta$
also leads to the same $B_{\rm tot}$ and $U_{\pm}/U_{\rm B}$. 
In the same way as shown above,
the maximum and minimum values of $B_{\rm tot}$
are determined by the jet power limit and minimum size limit
at the EHT-region.
However,  $U_{p}$ is much smaller than $U_{B}$
simply because of the paucity
of the relativistic proton component.

\subsubsection{The cases for non-relativistic protons ($\zeta=1$)}

Next, let us consider the case of  non-relativistic protons ($\zeta=1$).
When non-relativistic  protons are loaded,
the corresponding energetic condition can be given by
$L_{\rm jet} \ge 
\max
\left[L_{\rm poy}, 
\left(1+\frac{\eta}{2-\eta}\right)
L_{\pm}\right] $.
Since the protons are non-relativistic,
the effect of proton loading  is quite small in terms of energetics.
The coefficient  resides in a narrow range
$1<(1+\eta/(2-\eta))<3/2$.
Note that $RM$ strongly depends on $\eta$ 
while $RM$ is independent of $\zeta$.

The "$e^{-}/p$-dominated" case  results in  similar 
values of $B_{\rm tot}$ and $U_{\pm}/U_{\rm B}$
 to those shown in Table~\ref{table:eta99zetamemp}, 
because
the maximum and minimum values of $B_{\rm tot}$
are also 
determined by the jet power limit and minimum size limit
at the EHT-region.
The contribution of protons are only $U_{p}=U_{\pm}/2$. 
So, it does not give any significant effects on energetics.

Finally, we comment on 
the "$e^{\pm}$-dominated" case.
The main difference between the 
"$e^{-}/p$-dominated" and "$e^{\pm}$-dominated" cases is
$n_{e^{-}p}$.
Since the number density of $e^{\pm}$-pairs
does not contribute to $RM$, the constraint
of $RM$ becomes weaker when 
 $n_{e^{-}p}$ becomes smaller.
 It  leads to wider allowed region for smaller 
 $\gamma_{\pm}$ and smaller $B$ region.
 Therefore, the maximum value of allowed
 $U_{\pm}/U_{B}$ for  the "$e^{\pm}$-dominated" case
 becomes larger than that for "$e^{-}/p$-dominated" case.
 However, this only changes the allowed  $\gamma_{\pm}$
 within a factor of $\sim 10$ and it does not give a large 
 impact
 on energetics.

\section{Fully SSA-thin case}

It is worthwhile to examine a case of
fully SSA-thin model for EHT-region
since the indication of  $\nu_{\rm ssa}>100~{\rm GHz}$ by
interferometry observations does not necessarily
mean that $\nu_{\rm ssa}$ is larger than $230~{\rm GHz}$.
We can safely regard the SSA frequency as
$43~{\rm GHz} < \nu_{\rm ssa} <  230~{\rm GHz}$
where the lower limit is warranted by the detection of 
core-shift at 43~GHz in Hada et al. (2011).

In Figure~\ref{fig:spectrum}, we show a schematic draw
of the synchrotron spectrum when
the EHT-region is SSA-thin at 230~GHz (solid line).
The upper limit of the  flux density at 43~GHz 
of the 230~GHz core is estimated 
as $0.09~{\rm Jy}=0.7~{\rm Jy}\times (40/110)^{2}$
based on the VLBA measurements of the radio core 
flux and size by Hada et al. (2013). 
The gray-colored scale shows the typical flux density 
obtained by SMA and CARMA. 
Interferometric observation shows 
some variability at 230~GHz (Akiyama et al. 2015).
We define this as $F_{\rm upper}$
and we assume that  $F_{\rm upper}$ is the 
upper limit of the flux density in overall frequency range 
of $43~{\rm GHz} < \nu_{\rm ssa} <  230~{\rm GHz}$.
First, from the EHT data, we can estimate a possible
lower limit of $\nu_{\rm ssa}$ as
\begin{eqnarray}
\nu_{\rm ssa} &\ge& 230~{\rm GHz}\times 
\left(\frac{F_{\rm upper}/2.3~{\rm Jy}}{S_{\nu}/1~{\rm Jy}}\right) ^{-1/\alpha} \nonumber \\
&\sim &
160~{\rm GHz} ~ ({\rm for~\alpha=2.5} ) .
\end{eqnarray}
Note that, the choice of $\alpha=3.0$ leads to 
$\nu_{\rm ssa} \sim 170~{\rm GHz}$.
Second, from the VLBA data, we can estimate a possible
upper limit of $\nu_{\rm ssa}$ as
\begin{eqnarray}
\nu_{\rm ssa} &\le& 43~{\rm GHz}\times 
\left(\frac{F_{\rm upper}/2.3~{\rm Jy}}{S_{\nu}/0.09~{\rm Jy}}\right) ^{2/5} \nonumber \\
&\sim & 160~{\rm GHz} .
\end{eqnarray}
Allowing some flux measurement errors,
somehow we can have consistent case 
around $\nu_{\rm ssa}\sim 160~{\rm GHz}$ with $\alpha\sim 2.5$.

Then, let us discuss on physical quantities in this case.
From Eq.~\ref{eq:ueub},  
 $U_{\pm}/U_{B} \propto \nu_{\rm ssa}^{-2p-13}$. 
Therefore, in this case, 
the ratio would be typically larger by a factor of
$(160/230)^{-18}\sim 6.9\times 10^{2}$ (for $p=2.5$)
than that for
the SSA-thick case. 
However, 
this does not change the result of $U_{\pm}\ll U_{\rm B}$
since $U_{\pm}\ll U_{\rm B}<10^{-4}$ in any cases
with the SSA-thick core existing.
Hence, we can conclude that even for
fully SSA-thin EHT-region case, 
 $U_{\pm}\ll U_{\rm B}$ holds in order not to 
 overproduce fluxes between
 $43~{\rm GHz} < \nu_{\rm ssa} <  230~{\rm GHz}$.

However, a critical difference appears 
for the comparison between $U_{p}$ and $U_{\rm B}$.
In the case of relativistic protons with "$e^{-}/p$-dominated composition, 
$U_{p}>U_{B}$ can be realized for a certain range of $\nu_{\rm ssa}$.
From the Table~\ref{table:eta99zetamemp}
we know the values of 
$U_{p}/U_{B}$ when $\nu_{\rm ssa}=230~{\rm GHz}$.
By multiplying the factor of $\sim 200-400$ , 
and the maximum  value reaches $U_{p}/U_{B} >1$
at $\nu_{\rm ssa}\sim 160~{\rm GHz}$.

\section{Summary}

We have explored the magnetization degree of the jet base of M87 
based on the observational data  of
the EHT obtained by Doeleman et al. (2012).
Following the method in K14,
we estimate the energy densities of magnetic fields ($U_{B}$) and
electrons and positrons ($U_{\pm}$) in the region detected by EHT
(EHT-region) with its FWHM size 
$40~{\rm \mu as}$.
Imposing basic energetics of  the M87 jet,
the constraints from EHT observational data, 
and the minimum size of the SSA-thick region as 
the ISCO radius,
we find  the followings.

\begin{itemize}

\item

First, we adopt the assumption that 
the  EHT-region contains an SSA-thick region. 
Then, the co-existence of SSA-thick and SSA-thin regions
is required for the EHT-region not to overproduce 
$L_{\rm poy}$.
The angular size of the SSA-thick region is
limited as $21~{\rm \mu as} \le \theta_{\rm thick}\le 25.5~{\rm \mu as}$,
while that of the SSA-thin region should be $40~{\rm \mu as}$
to explain the EHT data.
The derived flux density of the SSA-thick region is about 0.27~Jy.
The allowed magnetic-fields strength in the SSA-thick region is 
$58~{\rm G}\le B_{\rm tot}\le127~{\rm G}$.
In terms of energetics, $U_{B}\gg U_{\pm}$
is realized at the overall SSA-thick region. 
If protons do not dominantly contribute to jet energetics,
then this result supports the magnetic-driven jet scenario 
at the SSA-thick region.

We further examine the following four cases
for electron/positron/proton ($e^{-}/e^{+}/p$) mixed plasma;
non-relativistic protons with $e^{-}/p$-dominated composition,
non-relativistic protons with $e^{\pm}$-dominated composition,
relativistic protons with $e^{-}/p$-dominated composition, and
relativistic protons with $e^{\pm}$-dominated composition, 
together with the assumption that 
$RM$ detected by SMA (Kuo et al. 2014) 
gives an upper limit of $RM$ of the EHT-region.
Although $RM$ limit can give tighter constraints
on allowed $\gamma_{\pm}$, 
it does not change the results significantly. 
We find that  $U_{B}\gg U_{\pm}$  
always holds in any case.

\item

Second, the case of 
completely SSA-thin ($\nu_{\rm ssa}< 230~{\rm GHz}$)
EHT-region is also discussed.
Although lower $\nu_{\rm ssa}$ can
increase the ratio $U_{\pm}/ U_{\rm B}$ by a factor of
$200-400$ than that for
the SSA-thick case,
this does not change the result of $U_{\pm}\ll U_{\rm B}$
since $U_{\pm}/U_{\rm B}<10^{-3}$.
However, we also find that,
in the case of relativistic protons with 
"$e^{-}/p$-dominated" composition, 
$U_{p}>U_{B}$ can be realized
around $\nu_{\rm ssa}\sim 160~{\rm GHz}$.

\end{itemize}

Future work and key questions  are enumerated below.

\begin{figure} 
\includegraphics[width=8cm]{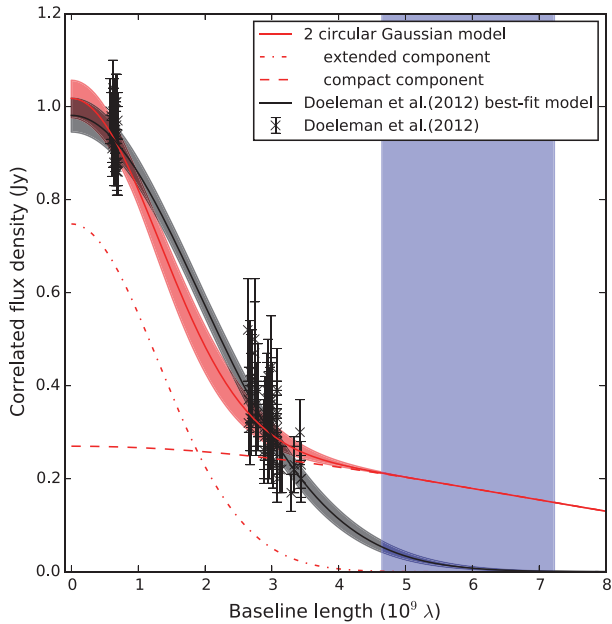}
\caption
{Gaussian fittings to the correlated flux density of the M87 core
obtained by EHT at 230~GHz.  
The flux density data plotted as a function
of baseline length are adopted from Doeleman et al. (2012). 
The black solid curve is the best-fit circular Gaussian
model with $S_{\nu}=0.98~{\rm Jy}$ and 
$\theta_{\rm FWHM}=40~{\rm \mu as}$
 obtained by Doeleman et al. (2012).
The red solid curve is the best-fit two-component model.
The red dashed  and dot-dashed curves represent
the SSA-thick and the SSA-thin components,
respectively.  
 The SSA-thick component
is expressed as the gaussian with  
$\theta_{\rm FWHM}=21~{\rm \mu as}/1.8=11.1~{\rm \mu as}$ 
and $S_{\nu}=0.27~{\rm Jy}$.
The size and the flux density of  the
extended SSA-thin component are
$\theta_{\rm FWHM}=60~{\rm \mu as}$ and
$S_{\nu}=0.75~{\rm Jy}$.
The blue-shaded region represents
the baseline-length range corresponding
to the one between the Hawaii/Arizona/California
and Chile.}
\label{fig:radplot}
\end{figure}

\begin{figure} 
\includegraphics[width=8cm]{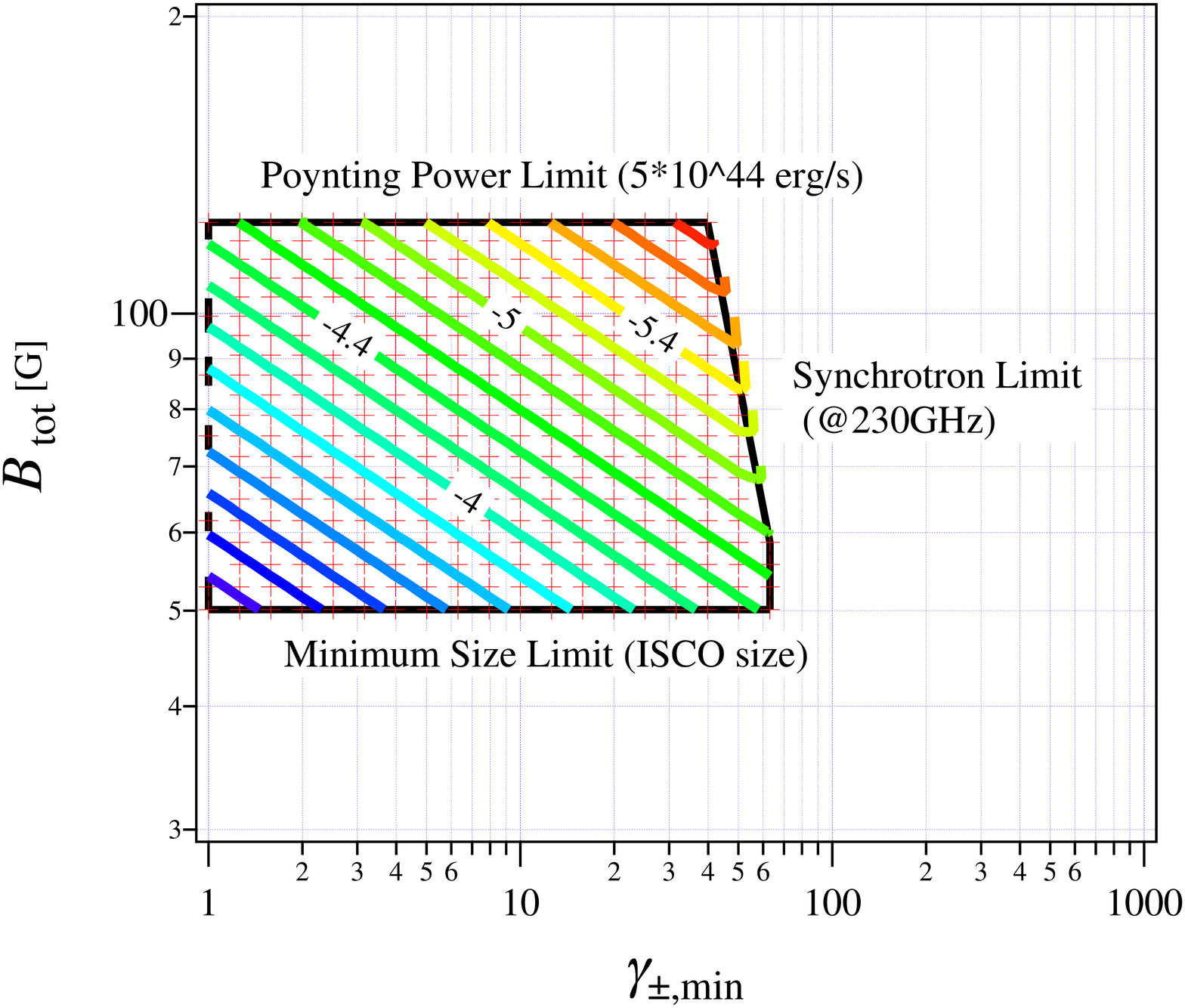}
\caption
{The allowed region of 
$\gamma_{\pm,\rm min}$ and $B_{\rm tot}$
(the red cross points enclosed by the black trapezoid).
The  colored contour lines show the  allowed
$\log (U_{\pm}/U_{\rm B})$. 
The tags $\log (U_{\pm}/U_{\rm B})=$-4, -4.4, -5, and -5.4
are marked as reference values. 
The physical quantities and parameters adopted are 
$L_{\rm jet} = 5 \times 10^{44}~{\rm erg s^{-1}}$, and 
$p=3.0$.
The minimum $\gamma_{\pm}$ is limited by $\nu_{\rm syn,\rm obs}$
at 230~GHz.}
\label{fig:synlimit_fiducial}
\end{figure}

\begin{figure} 
\includegraphics[width=8cm]{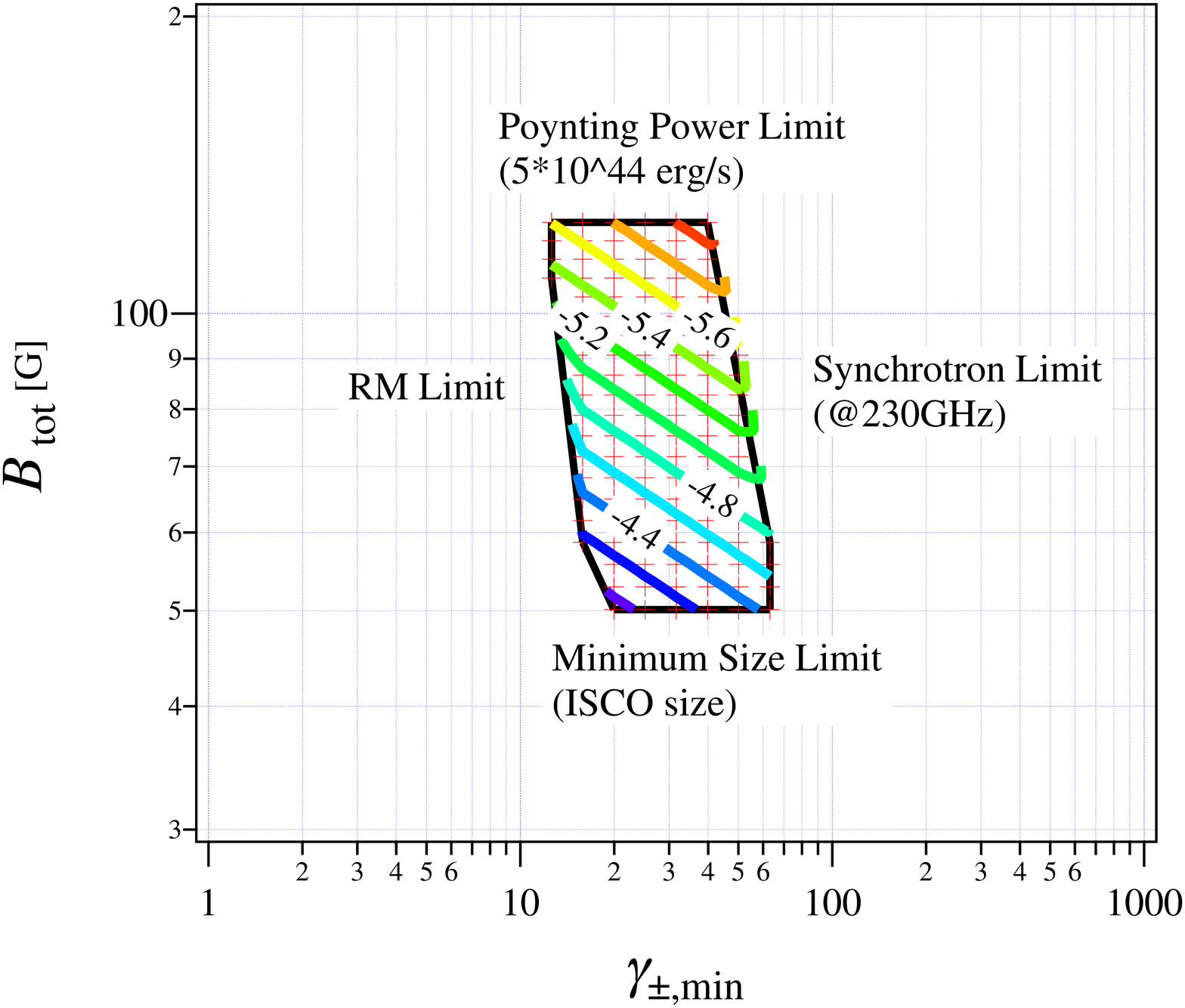}
\caption
{
The allowed region of 
$\gamma_{\pm,\rm min}$ and $B_{\rm tot}$
when $RM$ limit is taken into account.
The physical quantities and parameters adopted are 
$L_{\rm jet} = 5 \times 10^{44}~{\rm erg s^{-1}}$, 
$p=3.0$,
$\eta=0.99$ and 
$\zeta=m_{e}/m_{p}$ which
corresponds to 
$e^{-}/p$-dominated composition with
relativistic protons.
The tags $\log (U_{\pm}/U_{\rm B})=$ -4.4, -4.8, -5.2, -5.4, and -5.6
are marked as reference values.}
\label{fig:eta99zetamemp}
\end{figure}

\begin{figure} 
\includegraphics[width=8cm]{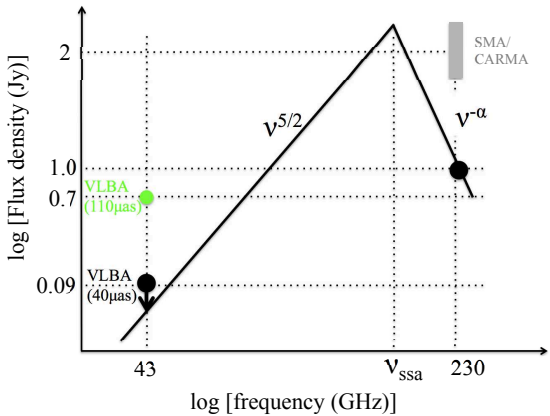}
\caption
{Schematic view of synchrotron spectrum when
the EHT-region is fully SSA-thin at 230~GHz 
with its size and flux density $40~{\rm \mu as}$ and 1.0~Jy (solid line).
The upper limit on the flux density at 43~GHz is estimated 
as $0.09~{\rm Jy}=0.7~{\rm Jy}\times (40/110)^{2}$
based on the VLBA measurements of core size and flux at 43~GHz
by Hada et al. (2013). 
The gray-colored range shows the typical
flux density  at 230~GHz obtained by SMA and CARMA
(e.g., Doeleman et al. 2012; Kuo et al. 2014; Akiyama et al. 2015).}
\label{fig:spectrum}
\end{figure}

\begin{itemize}

\item

An important future work is 
to confirm the existence of  the SSA-thick region in the EHT-region.
If we confirm it,
then we can exclude the case of $U_{p}/U_{B}>1$.
In the context of confirming the existence of SSA-thick region,
we also add to note the effectiveness of 
inclusions of longer baselines even for a single frequency VLBI observation.
In Fig~\ref{fig:radplot},
it is clear that the visibility amplitude of the SSA-thin component
is much smaller than that of SSA-thick component 
above  $\sim 3G\lambda$  at 1.3mm wavelength. 
Therefore, inclusions of baselines with $>3G\lambda$
 would be effective to distinguish the SSA-thick component.
For example, phased ALMA plus SMT with an effective bandwidth
of 4~GHz would be effective at  $\sim 5G\lambda$ 
(Fig~6 in Fish et al. 2013).
In Figure~\ref{fig:radplot}, we show the 
corresponding baseline-length range (the blue-shaded region).

\item

Equally important future work is 
to observe the EHT-region with the spatial resolution of
$\sim 1~R_{s}$ of M87. 
Currently, the  EHT array with 20-30~$\mu$as 
resolution at 230 and 345~GHz (e.g., Lu et al. 2014)
is not able to reach   $\sim 1~R_{s}$ of M87. 
Ground-based short-mm VLBI observations are very sensitive to
weather conditions (e.g., Thompson et al. 2001).
To confirm our assumption that the minimum $\theta_{\rm thick}D_{L}$ 
is comparable to $\sim R_{\rm ISCO}$ or even smaller,
space VLBI observations would be required in future.
In the past missions and existing plan of space VLBI,   
it was not able to reach the event horizon scale of M87
(e.g., Dodson et al. 2006; Asada et al. 2009; Dodson et al. 2013)
since target wavelength were not short enough.
Thus, atmospheric-free space (sub-)mm VLBI observation would 
be indispensable to reach $\sim 1~R_{s}$ of M87.
The phased ALMA (e.g., Alef et al. 2013, Fish et al. 2013)
will  play a definitive role for such observations for obtaining 
visibilities between space and ground telescopes baselines.

Honma et al. (2014) have recently
proposed a new technique of VLBI data-analysis
to obtain super-resolution images with radio interferometry
using sparse modeling.
The usage of the sparse modeling enables us to obtain 
super-resolution images in which structure finer than
the standard beam size can be recognized. 
A test simulation for imaging of the jet base of M87 is
actually demonstrated in  Honma et al. (2014) and 
the technique works well.
Therefore, this super-resolution technique
will become another important tool
for obtaining better resolution images.

\item
The observational result of 
Doeleman et al. (2012) does not show flux
variability at 230~GHz.
However, total epoch-number of EHT observations
is  too scarce to confirm the absence of flux variability 
at 230~GHz all of the time.
M87 might be in  quiescent state  during the EHT observations
in April 2010 by chance.
We also emphasize that 
the derived field strength is still $\ge 58~{\rm G}$
and $t_{\pm,\rm syn}$ still tends to be smaller than day scale.
It is also intriguing that the same correlated flux densities 
in 2009 reported by Doeleman et al. (2012)
are observed during another EHT observation performed
in April 2012 (Akiyama et al. 2015).
This result is quite different from the day scale variability
detected in Sagittarius A*  by the EHT observations
(Fish et al. 2011).
To search for a possible flux variability of M87
in more details, continuous monitoring by EHT
 would be essential.

\item

Based on GRMHD model,
well-ordered poloidal fields are dominant within the Alfven point and 
toroidal fields become dominant outside of Alfven point
while turbulence may not grow-up yet at the jet base
(e.g., Spruit 2010 for review).
In general, turbulent eddies which most probably generate
turbulent fields are not expected before  sufficient
interactions with surrounding ambient matter
(e.g., Mizuta et al. 2010 and reference therein).
Therefore, higher LP degree is likely to be expected.
Conservatively saying,
the reason of low LP degree  by Kuo et al. (2014)
is most probably because of depolarization
within SMA beam.
At the moment, we are not able to rule out
a possible constitution of RIAF emission which may also
lead to low LP degree.
If so, then studies of fundamental process for 
particle accelerations in RIAF (e.g., Hoshino 2013) and
 the effects particle escape from RIAF
 (Le \& Becker 2004;
 Toma \& Takahara 2012; 
 Kimura et al. 2014)
 would become more important.

\item

In terms of 
the brightness temperature of the 230~GHz radio core of M87
$T_{b} \sim 2 \times 10^{10}~{\rm K}
\left(\frac{S_{\nu}}{1~{\rm Jy}}\right)
\left(\frac{\theta_{\rm FWHM}}{40~{\rm \mu as}}\right)^{-2}$
seems slightly higher than 
the prediction of hot electron temperature of $\sim 10^{9}~{\rm K}$
in RIAF flows (e.g., Manmoto et al. 1997).
Hence, the jet emission seems to be preferred to explain 
EHT-emission in M87 (Dexter et al. (2012),
see Ulvestad \& Ho (2001) for  similar arguments).
However, it is not conclusive because geometry
near ISCO regions is  highly uncertain in observational point of view.
The scrutiny of the origin of the 230~GHz emission
is still a noteworthy big issue to explore.

\item
Further　polarimetric observation would be required to 
examine RM properties in more details.
Although we adopt $RM$ values of Kuo at al. (2014),
it is found that the observed 
electric vector position angle (EVPA) trend does not show
a sufficiently tight fit to $\lambda^{2}$-law. 
This behavior may not be due to the consequence of 
blending of multiple cross-polarized
sub-structures with different $RM$ values, 
but simply rather due to the non-uniformity
between the upper and lower side bands of the SMA.
A polarimetric observation with ALMA is clearly 
one of the promising first step to improve this point.
Obviously, in the final stage,
short-mm (and sub-mm) VLBI polarimetric observations are inevitable
to avoid the contamination from the extended region.

\item 
Degree of $e^{\pm}/p$ coupling is a critical factor 
for the results of the proton power.
Theoretically, Hoshino and Arons (1991)  found 
the energy transfer  process from 
protons to positrons via absorptions of 
high harmonic ion cyclotron waves emitted by the protons.
Amato and Arons (2006) indeed performed 
one-dimensional particle-in-cell (PIC) 
simulations for  $e^{-}/e^{+}/p$-mixed plasma.
However, there are several simplifications in PIC
simulations such as smaller $m_{p}/m_{e}$ ratio etc.
More intensive investigations are awaited to 
clarify the degree of $e^{\pm}/p$　coupling at the base of the M87 jet.

\item 

We make a brief comment on 
effects of magnetic field topology and 
anisotropy of $e^{-}/e^{+}$ in terms of 
energy distribution.
If $e^{-}/e^{+}$ energy distribution in the EHT region is isotropic, 
then the synchrotron absorption coefficient investigated by GS65
is applicable and  differences of field-geometry
would not have an impact on field strength estimation.
For example,  the difference of  $B_{\rm tot}$ between
the cases of isotropic field (see Eq. (1)) and
ordered field ($B_{\perp}=B_{\rm tot}$) which is directed towards
LOS is only a factor of $\sqrt{3/2}$.

However, if the $e^{-}/e^{+}$ energy distribution is 
highly anisotropic, 
then the well known synchrotron emissivity and
self-absorption coefficient are not applicable.
Effects of the $e^{-}/e^{+}$ anisotropy on synchrotron radiation
are not well studied and it is beyond the scope of this paper.
Although we do not have any observational suggestions of  
such anisotropy of  $e^{-}/e^{+}$ energy distribution, 
it may be a new theoretical topic to be explored if observational 
suggestions are found in the future.

\end{itemize}


\bigskip
\leftline{\bf \large Acknowledgments}
\medskip

\noindent

We thank the anonymous referee for constructive comments.
KH and KA are
supported by the Japan Society for the Promotion
of Science (JSPS) Research Fellowship Program for Young
Scientists.


\begin{appendix}

\section{Modification of numerical factors}

In order to obtain better accuracy  calculation plus some modifications
of the definition of $B_{\perp}$ and relevant corrections, 
modified numerical co-efficient of $b(p)$ and $k(p)$ are
presented although the corrections are small.

In K14, magnetic fields strength perpendicular to the local
electron motions were not averaged over the pitch angle
(In Eqation~(1) in K14). 
In this work, in Eqation~(\ref{eq:Btot-define}),
we conduct the pitch-angle averaging for defining
the averaged magnetic fields strength perpendicular to the local
electron motions.

Synchrotron self-absorption coefficient measured
in the comoving frame is given by
(Eqs.~4.18 and 4.19 in GS65; Eq. 6.53 in Rybicki \& Lightman 1979)
\begin{eqnarray}
\alpha_{\nu}=
\frac{\sqrt{3}e^{3}}{8\pi m_{e}}
\left(\frac{3e}{2\pi m_{e}^{3}c^{5}}\right)^{p/2}
c_{1}(p) 
K_{\pm} B_{\perp}^{(p+2)/2}
\nu^{-(p+4)/2},
\end{eqnarray}
where the numerical coefficient $c_{1}(p)$
is expressed by using the gamma-functions as follows;
$c_{1}(p)=
\Gamma[(3p+2)/12]
\Gamma[(3p+22)/12]$.
For convenience, we define 
$\alpha_{\nu}=X_{1}c_{1}(p)B_{\perp}^{(p+2)/2}K_{\pm}\nu^{-(p+4)/2}$.

Optically thin synchrotron emissivity
per unit frequency $\epsilon_{\nu}$ 
from uniform emitting region is given by
(Eqs. 4.59 and 4.60 in BG70;
see also Eqs. 3.28, 3.31 and 3.32 in GS65)
%
\begin{eqnarray}
\epsilon_{\nu} =
4\pi\frac{\sqrt{3}e^{3}}{8 \sqrt{\pi} m_{e}c^{2}}
\left(\frac{3e}{2\pi  m_{e}^{3}c^{5}}\right)^{(p-1)/2} 
c_{2}(p)  K_{\pm}B_{\rm tot}^{(p+1)/2} \nu^{-(p-1)/2}   ,
\end{eqnarray}
where the numerical coefficient is 
$c_{2}(p)=
\Gamma[(3p+19)/12]
\Gamma[(3p-1)/12)]
\Gamma[(p+5)/4)]/\Gamma[(p+7)/4)]
/(p+1)$.
In K14,  $B_{\rm tot}$ was wrongly written
as $B_{\perp}$. So, here we revise it and it leads to
larger $b(p)$ by the factor of $\sqrt{1.5}$.
For convenience, we define 
$\epsilon_{\nu} \equiv 4\pi X_{2} c_{2}(p) B_{\rm tot}^{(p+1)/2}K_{\pm}
\nu^{-(p-1)/2}$.
The  modified coefficient $b(p)$ is expressed as
\begin{eqnarray}
b(p)=
\left(\frac{4\pi c_{2} X_{2} \times 1.5^{1/4}}{6 c_{1}X_{1}}\right)^{2} 
\times 2^{-4} .
\end{eqnarray}
In K14, the index of square bracket at the right hand side of 
$b(k)$ should not be $2$ but $-2$　(typo).
The expression of $k(p)\propto b(p)^{(-p-2)/2}$ does not change,
but the value $k(p)$ is changed.
Although the modifications 
of $b$ and $k$ in Table 1 of K14 are straightforward based on 
the above explanations, we put the table \ref{table:b-k} for convenience.

 \end{appendix}



\newpage


\newpage

\input{table1}
\input{table2}
\input{table3}

\end{document}

%% file: table1.tex
\begin{table*}
\centering
\caption{Results when the EHT region contains SSA-thick region}
\label{table:synlimit_fiducial}       
\begin{tabular}{lcccc}
\hline\noalign{\smallskip}
{\bf $L_{\rm j}$}    & 
allowed {\bf $B_{\rm tot}$} &  
allowed {\bf $\theta_{\rm thick}$} & 
allowed $U_{\pm}/U_{B}$& 
\\

[erg~s$^{-1}$] &
[G] &
[$\mu$as] &
                    &
\\
\noalign{\smallskip}\hline\noalign{\smallskip}


$5\times 10^{44}$ &
$50 \le B_{\rm tot}\le 124$&
$21 \le \theta_{\rm thick} \le 26.3$ &
$ 7.9\times 10^{-7} \le \frac{U_{\pm}}{U_{B}}\le 2.3\times 10^{-3}$&
\\

\noalign{\smallskip}\hline
\end{tabular}

\end{table*}

%% file: table2.tex
\begin{table*}
\centering
\caption{Results for the case of $e^{-}/p$-dominated composition with relativistic protons}
\label{table:eta99zetamemp}       
\begin{tabular}{lccccccc}
\hline\noalign{\smallskip}
{\bf $\eta$}    & 
$e^{+}$ fraction & 
{\bf $L_{\rm j}$}    & 
allowed {\bf $B_{\rm tot}$} &  
allowed  {\bf $\theta_{\rm thick}$} & 
allowed $U_{\pm}/U_{B}$& 
allowed $U_{p}/U_{B}$& 
\\
  &
[\%] &
[erg~s$^{-1}$] &
[G] &
[$\mu$as] &
&
\\
\noalign{\smallskip}\hline\noalign{\smallskip}

0.9 &
10&
$5\times 10^{44}$ &
$50\le B_{\rm tot}\le 124 $&
$ 21\le\theta_{\rm thick}\le 26.3$&
$ 7.9\times 10^{-7} \le \frac{U_{\pm}}{U_{B}} \le 1.1\times 10^{-4}$ &
$ 1.2\times 10^{-3} \le \frac{U_{p}}{U_{B}} \le  0.17$ &
\\

0.99 &
1&
$5\times 10^{44}$ &
$50\le B_{\rm tot}\le 124 $&
$ 21 \le\theta_{\rm thick}\le 26.3$&
$ 7.9\times 10^{-7} \le \frac{U_{\pm}}{U_{B}}\le 1.1\times 10^{-4}$ &
$ 1.4\times 10^{-3}\le \frac{U_{p}}{U_{B}} \le 0.20$ &
\\

1 &
0 &
$5\times 10^{44}$ &
$50\le B_{\rm tot}\le 124 $&
$ 21 \le\theta_{\rm thick}\le 26.3$&
$ 7.9\times 10^{-7} \le \frac{U_{\pm}}{U_{B}}\le 1.1\times 10^{-4}$ &
$ 1.4\times 10^{-3}\le \frac{U_{p}}{U_{B}}\le 0.21$ &
\\

\noalign{\smallskip}\hline
\end{tabular}

\end{table*}

%% file: table3.tex
\begin{table*}
\centering
\caption{Relevant coefficients for $B_{\perp}$ and $K_{\pm}$}
\label{table:b-k}       
\begin{tabular}{lccccc}
\hline\noalign{\smallskip}
{\bf $p$}    & 
{\bf $b(p)$} & 
{\bf $b(p)$} in K14 & 
{\bf $b(p)$} in Hirotani (2005)&
{\bf $b(p)$} in Marscher (1983)&
{\bf $k(p)$} 
\\
\noalign{\smallskip}\hline\noalign{\smallskip}

 2.5 &
 $4.1\times 10^{-5}$ &
$3.3\times 10^{-5}$ &
$2.36\times 10^{-5}$ &
$3.6 \times 10^{-5}$ &
$9.3\times 10^{-3}$ 
\\

3.0 &
 $2.4\times 10^{-5}$ &
$1.9\times 10^{-5}$ &
$2.08\times 10^{-5}$ &
$3.8\times 10^{-5}$ &
$1.4\times 10^{-3}$ 
\\

3.5 &
$1.5\times 10^{-5}$ &
$1.2\times 10^{-5}$ &
$1.78\times 10^{-5}$ &
-- &
$2.1\times 10^{-4}$ 
\\

\noalign{\smallskip}\hline
\end{tabular}

\end{table*}